\documentclass[twocolumn]{aastex63}
\usepackage{amsmath}

\usepackage{import}
\usepackage{hyperref}
\setcitestyle{authoryear}
\usepackage{scalerel}

\newcommand*{\kms}{\ensuremath{\mathrm{km}\,\mathrm{s}^{-1}}\,}

\shorttitle{Forecasting uncertainties}
\shortauthors{Guerra et al.}
\graphicspath{{./}{figures/}}

\begin{document}
    \title{Forecasts on the Dark Matter Density Profiles of Dwarf Spheroidal Galaxies with Current and Future Kinematic Observations}

    \author{Juan Guerra}
    \affiliation{Department of Astronomy, Yale University, New Haven, CT 06520, USA}

    \author{Marla Geha}
    \affiliation{Department of Astronomy, Yale University, New Haven, CT 06520, USA}

    \author{Louis E. Strigari}
    \affiliation{Mitchell Institute for Fundamental Physics and Astronomy, Department of Physics and Astronomy, Texas A\&M University, College Station, TX 77845, USA}

\begin{abstract}
    We forecast parameter uncertainties on the mass profile of a typical Milky Way dwarf spheroidal (dSph) galaxy using the spherical Jeans Equation and Fisher matrix formalism.
    We show that radial velocity measurements for 1000 individual stars can constrain the mass contained within the effective radius of a dSph to within $5\%$.
    This is consistent with constraints extracted from current observational data.
    We demonstrate that a minimum sample of 100,000 (10,000) stars with both radial and proper motions measurements is required to distinguish between a cusped or cored inner slope at the 2-sigma (1-sigma) level.
    If using the log-slope measured at the half-light radius as a proxy for differentiating between a core or cusp slope, only 1000 line-of-sight and proper motions measurements are required, however, we show this choice of radius does not always unambiguously differentiate between core and cusped profiles.
    Once observational errors are below half the value of the intrinsic dispersion, improving the observational precision yields little change in the density profile uncertainties.
    The choice of priors in our profile shape analysis plays a crucial role when the number of stars in a system is less than 100, but does not affect the resulting uncertainties for larger kinematic samples.
    Our predicted 2D confidence regions agree well with those from a full likelihood analysis run on a mock kinematic dataset taken from the Gaia Challenge, validating our Fisher predictions.
    Our methodology is flexible, allowing us to predict density profile uncertainties for a wide range of current and future kinematic datasets.
\end{abstract}

\section{Introduction}\label{sec:intro}

    The motions of individual stars provide a powerful tracer of the underlying mass distribution in galaxies.
    For dark matter-dominated galaxies, measurements of the enclosed central mass or the inner density slope can differentiate between dark matter models.
    Cold dark matter (CDM) predicts cuspy central density profiles~\citep{Navarro2010-dy}, while cored central density profiles~\citep{Burkert1995-gc} may be generated in models such as self-interacting dark matter~\citep{Spergel:1999mh}, warm dark matter~\citep[][]{Newton2021-ph}, fuzzy dark matter~\citep{Hu:2000ke}, as well as due to baryonic processes \citep{Governato2015}.
    Measurements of the dark matter distributions may also be used to place constraints on the dark matter particle annihilation cross section~\citep[e.g.,][]{Strigari2018-fz}.

    The dwarf satellite galaxies around the Milky Way \citep{Mateo1998-ai, McConnachie2012-ta, Simon2019-xg} are sufficiently nearby to resolve individual stars, maximizing available kinematic information.
    Although in the brightest of dwarf spheroidal (dSph), baryonic process can change the shape of the inner-slope~\citep[e.g.,][]{Governato2012-cc}, the low mass end of dSphs and ultra-faint (UFDs) satellites have sufficiently low stellar masses to mitigate concern that baryonic processes have modified their underlying density profiles~\citep[e.g,][]{Penarrubia2012-uj,Di_Cintio2014-mt}.
    The dSph and UFDs contain old stellar populations \citep{Brown2014-tu} and are largely devoid of gas \citep{Putman2021-ab}.
    Radial velocities are typically available for a few dozen stars in UFD systems \citep[e.g.,][]{simon2007}, while the more luminous dSphs have hundreds to a thousand individual radial velocity measurements~\citep[e.g.,][]{Walker2009-lm}.
    Ensemble proper motions are measured for many Milky Way satellites \citep{2018A&A...616A..12G,simon2019-ba,Fritz2018-ft,mcconnachie2020}, however, the number of individual stars with transverse (proper) motions in these systems remains small \citep{Massari2020-ue}.
    This paper is motivated by the near-term prospect of increased proper motion measurements of individual stars in Milky Way satellite galaxies.

    Given that dSphs are dark matter dominated and are amongst the few systems for which large samples of individual stars may be resolved, they are an ideal laboratory to study the nature of dark matter  (for a review see~\citealt{Battaglia2013-we,Simon2021-ax}).
    The most common dynamical modeling method using individual stars as tracers of the gravitational potential is spherical Jeans modeling (e.g.,~\citealt{Jeans1915-an,Binney2011-ja,Mamon2013-bi,Cappellari2015-jt,Read2017-nl}).
    More sophisticated methods are also used to study these systems such as phase-space methods~\citep[e.g.,][]{Wojtak2010-im,Strigari:2018bcn}, made-to-measure~\citep[e.g.,][]{Williams2015-gl}, action-angle modeling~\citep[e.g.,][]{pascale2019} and Schwarzschild modeling~\citep[e.g.,][]{Schwarzschild1979-ni,Breddels2013-mn,Hagen2019-du}.

    While a variety of modeling methods have been used to study the kinematics of dSphs, the nature of their central dark matter densities remain inconclusive.
    Some studies indicate that dSphs are consistent with core profiles (e.g., \citealt{Gilmore2007-aa,Walker2011-og,Pascale2018-ud}), others favor NFW cusp profiles \citep[e.g.,][]{Strigari:2010un,Jardel2013-yj,Penarrubia2012-uj, Hayashi2020-cc}.
    The primary difficultly in extracting the central dark matter densities can be attributed to the degeneracy between the density profile, $\rho(r)$, and the stellar anisotropy, $\beta(r)$~\citep{Lokas2002-lk}.
    
    A proposed solution to the address the $\rho-\beta$ degeneracy is to measure the proper motions of individual stars~\citep{Wilkinson2002-vd,Strigari2007-cc,Hayashi2020-cc,Read2021-du}.
    Proper motions provide two additional velocity measurements, and when combined with radial velocity, allow for a full 3D reconstruction of the velocity vector.
    The addition of proper motions improves measurement of the velocity anisotropy, and thereby the central dark matter density.
    
    With the launch of the {\it Gaia} satellite \citep{2018A&A...616A..12G}, we are entering an era in which measurements of individual stellar proper motions in dSphs is becoming a reality.
    \citet{Massari2020-ue} combined previous epoch data from the {\it Hubble Space Telescope} with {\it Gaia} Data Release 2 results to measure the proper motions of two dozen stars in the Sculptor and Draco dSphs.
    Though these measurements represent an important achievement in kinematic studies of dSphs, the data sets are still too small to conclusively distinguish between NFW and cored dark matter halos \citep{Strigari:2018bcn}.
    Larger samples of proper motions will be achievable with the next generation of ground-based telescopes \citep[e.g.,][]{Evslin2015-ze,simon2019-ba} in combination with wide-area space-based missions \citep{Kallivayalil2020}.

    In this paper we study the prospects for precisely measuring the dark matter density profiles of dSphs, projecting what will be possible with current and future measurements of radial velocities and proper motions.
    While recent work has focused on using mock observations to explore similar questions \citep[e.g.,][]{Chang2021-bu, Read2021-du}, this paper extends the work of \citet{Strigari2007-cc}, using information theory, specifically the Fisher Information Matrix formalism, to predict uncertainties on important model parameters.
    Fisher formalism allows us to forecast parameter uncertainties without running a full computational-intensive likelihood analysis.
    We make our code fully available on GitHub at  \href{https://github.com/dmForecast/dmForecast}{dmForecast}\footnote{\url{https://github.com/dmForecast/dmForecast}}.

    The paper is organized as follows.
    In Section~\ref{sec:dynamical_modeling} we review the spherical Jeans Equation and how observable properties are related to the model parameters.
    In Section~\ref{sec:fisher} we introduce the Fisher matrix and how it is used to forecast uncertainties on model parameters.
    In Section~\ref{sec:analysis} we explore what information may be obtained from radial velocities and proper motions on the enclosed mass.
    We then consider the role that sample size, priors and velocity errors play on our predicted uncertainties.
    In Section~\ref{sec:comparisons} we specifically ask what type of observations are required to distinguish between a cusp and a core inner slope.
    In Section~\ref{sec:validattion} we validate our results by comparing our predicted uncertainties to those from a full likelihood analysis using mock data.
    Finally, in Section~\ref{conclusions}, we consider the broader applications of this method for optimizing future observations.

\section{Dynamical modeling}\label{sec:dynamical_modeling}
    The analysis in this work involves predicting uncertainties on mass profile parameters inferred from kinematic datasets.
    This requires that we explicitly understand how observations are related to the parameters of interest.
    We  first review the relevant equations (Section~\ref{ssub:jeans_equation}), parameters (Section~\ref{ssub:parameters}) and how these relate to observable quantities (Section~\ref{ssub:solution_jeans_equation}).

    \subsection{The Jeans Equation}\label{ssub:jeans_equation}
        The motions of stars within dwarf galaxies are well modeled as a collisionless fluid.
        That is, to a very good approximation their positions $\vec{x}$ and velocities $\vec{v}$ can be described by a distribution function, $f(\vec{x},\vec{v})$, that obeys the Collisionless Boltzmann Equation (CBE)\citep{Binney2011-ja}.
        This is true even in cases where mild tidal interactions are present \citep{Penarrubia2012-uj}, as is the case for the Milky Way's satellite dwarf galaxies.
        However, the distribution function can be a complex function of energy and angular momentum, and is usually poorly sampled even for Milky Way satellites.
        Thus, directly solving the CBE is challenging.
        A common method of modeling these systems consists of studying the statistical moments of the CBE.
        In particular, the second moment of the CBE results in the well known Jeans Equations\citep{Jeans1915-an,Binney2011-ja}.

        Assuming that a system is static, phased mixed, and spherically symmetric, there is only one non-trivial Jeans equation which we must solve in order to determine the dark matter potential \citep{Binney1982-ww,Binney2011-ja}:
        
        \begin{equation}\label{eq:spherical_jeans_eq}
            \frac{d(\nu_{\star} {\sigma_{r}^{2}})}{d r}
            + \frac{2\beta}{r}\nu_{\star}\sigma_{r}^{2}
                =-\nu_{\star}\frac{d \Phi_{\mathrm{tot}}}{d r}.
        \end{equation}
        Here $\nu_{\star}(r)$ is the three dimensional stellar density profile; $\sigma_r(r)$ is the radial velocity dispersion; $\Phi_{\mathrm{tot}}=\Phi_{\mathrm{dm}}+\Phi_{\star}$ is the gravitational potential profile due to both the dark matter and stellar component, and $\beta(r)$ is the stellar anisotropy profile.

    \subsection{Parameters}\label{ssub:parameters}
        In Equation~\ref{eq:spherical_jeans_eq}, $\nu_{\star}(r)$ is obtained from inverting the observed 2D stellar luminosity density $I(R)$ via the Abel transform \citep{Binney2011-ja}.
        In our analysis we will take $\nu_{\star}(r)$ to be a Plummer distribution (i.e we will take R to be random numbers drawn from this distribution).

        This leaves two unknown functions that must be inferred from kinematic observations: the underlying dark matter distribution and the stellar anisotropy profile.
        We choose parametric forms to describe both quantities.

        For the dark matter density distribution we chose the general double power law that is widely used to describe halos in cosmological simulations \citep{Zhao1996-vf,An2012-ji}:
        \begin{equation}\label{eq:density}
            \rho(r) = 
            \frac{\rho_{0}}{\left(\frac{r}{r_0}\right)^{a}\left[1+\left(\frac{r}{r_{0}}\right)^b\right]^{(c-a)/b}}
        \end{equation}
        where $a$ is the logarithmic slope at small radii, $c$ is the slope at large radii and $1/b$ sets the width of the transition region around the characteristic radius $r_{0}$.
        This five parameter (\{$\rho_0,r_0,a,b,c,$\}) model is extremely flexibility and reduces to many well known cases such as the Navarro-Frenk-White (NFW) profile $(a=1,b=1,c=3)$, Plummer profile $(a=0,b=2,c=5)$, or the generalized NFW profile, gNFW $(b=1,c=3)$.
        The only restriction is that one should be careful when choosing specific parameters since not every combination produces a positive definite distribution function \citep{Baes2021-kk}

        \begin{figure*}[htp]
            \includegraphics[width=.99\linewidth]{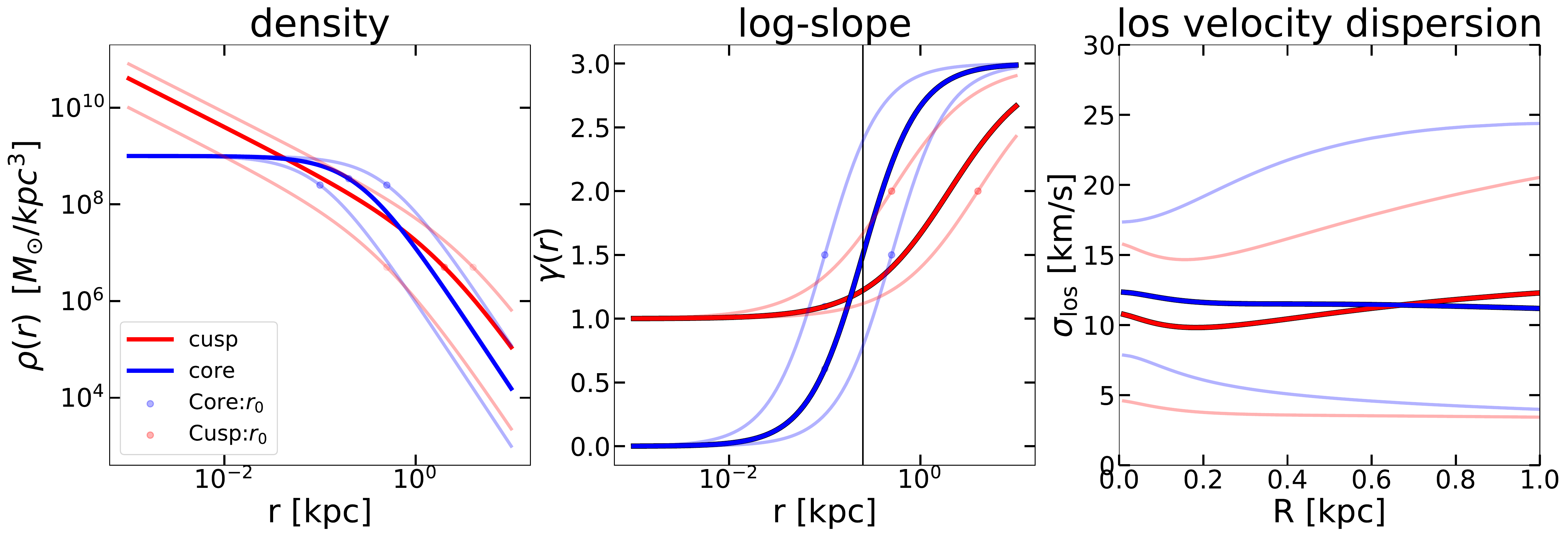}
            \caption{The density profile (left panel) for an NFW density profile (red) and a cored density profile(blue).
            We plot the corresponding log-slope curves (middle panel) and velocity dispersion profiles (right panel).
            The solid red/blue lines correspond to the fiducial cusp/core values used in our analysis.
            Faded curves show the same profiles as we vary the scale radius ($\rho(r_0)$).
            Circles indicate $\rho(r_0)$ and $\gamma(r_0)$ for these variations in the left and middle panels, respectively.
            The black vertical line in the middle panel indicates the effective radius ($r_{1/2}$) where the log-slope is typically measured.}
            \label{fig:densityvlogslope}
        \end{figure*}
        
        A useful quantity derived from density profile is the so-called log-slope, $\gamma(r) = \frac{\partial \ln\rho}{\partial\ln r}$.
        With our parametrization of the density function the log-slope behaves such that $\gamma(r\rightarrow 0)= a$ and $\gamma(r\rightarrow \infty)= c$.
        The shape of the curve in between these limits is dictated by scale radius $r_0$ and slope parameter $b$.
        At the half-light radius, the log-slope has been shown to be well constrained and can be used as an analog for the inner-slope~\citep{Strigari2007-cc,An2009-zo,Van_der_Marel2010-en,Read2018-bv}.
        Figure~\ref{fig:densityvlogslope} shows the fiducial density profiles we use in our analysis, as well as the corresponding log-slope and projected velocity dispersion profiles.
        We show the profiles for a variety of scale radii ($r_{0}$) to display how changing this parameter affects the behavior of the log-slope.
        The log-slope is commonly measured at the effective half-light radius which we define as $r_{1/2}$ (vertical line in the middle panel of Figure~\ref{fig:densityvlogslope}).

        In collisionless systems the stellar radial and tangential dispersions, $\sigma_r$ and $\sigma_t$, act as pressures that keep the system from collapsing.
        The stellar anisotropy, $\beta(r)$, is a measure of the local ratio of the tangential to radial dispersion:
        \begin{equation}\label{eq:stellar_anisotropy}
            \beta(r)
            = 1 -\frac{\sigma^2_t(r)}{\sigma^2_r(r)}.
        \end{equation}
        While we explore varying the value of $\beta$ in Section~\ref{ssub:mass}, for simplicity, we choose a fiducial value for the stellar anisotropy as $\beta = 0$ throughout the remainder of our analysis.

    \subsection{Solution to the Jeans Equation}\label{ssub:solution_jeans_equation}
        The solution to the spherical Jeans Equation can be found using ordinary differential equation techniques:
        \begin{equation}\label{eq:sigma_r}
            \nu_{\star}(r) \sigma_{r}^{2}(r)=
            \int_{r}^{\infty} \mathrm{d} r^{\prime}\nu_{\star}\left(r^{\prime}\right) \frac{\mathrm{d} \Phi}{\mathrm{d} r^{\prime}}\exp\left[2\int_{r}^{r'}\frac{\beta(u)}{u}du\right]
        \end{equation}
        
        In the case where only radial velocities are available, the dispersion above ($\sigma_{r}^{2}(r)$) is related to the observable line-of-sight (los) velocity dispersion ($\sigma_{\mathrm{los}}^{2}(R)$) via the Abel transform shown in Equation~\ref{eq:sigma_los}.

        With stellar proper motions in addition to line-of-sight velocities, full 3D information is available on the velocity vector.
        To relate this additional information to the model parameters we define a coordinate system.
        Consider a star at position $\vec{r}=(x,y,z)$ with a velocity vector $\vec{v} =v_x\hat{e}_x + v_y\hat{e}_{y}+v_z\hat{e}_z$.
        Assuming spherical symmetry, the velocity vector can be expressed as $\vec{v} =v_r\hat{e}_r + v_{\theta}\hat{e}_{\theta} +v_{\phi}\hat{e}_{\phi}$.
        To convert to observed velocities, we define a cylindrical coordinate system as $\vec{v}=v_{\rho}\hat{e}_{\rho}+v_{\phi}\hat{e}_{\phi}+v_{z}\hat{e}_{z}$.
        Here we take the $\hat{e}_{z}$ as the line-of-sight (los).
        We can then relate it to the spherical velocity components using $v_{i}=\vec{v}\cdot\hat{e}_{i}$ giving us.
        \begin{eqnarray}\label{eq:cylindrical_velocities}
            v_{\rho} &=& v_{r}\cos{\theta} - v_{\theta}\sin{\theta}\\
            v_{z} &=& v_{r}\sin{\theta} + v_{\theta}\cos{\theta}\\
            v_{\phi} &=& v_{\phi}.
        \end{eqnarray}
        We can now calculate the dispersions associated with these velocities noting that under our spherical symmetry assumptions $\sigma_i^{2} = \overline{v_{i}^{2}}$, $\overline{v_{r}v_{\theta}}=0$ and $\overline{v_{\phi}^2}= \overline{v_{\theta}^{2}}=\overline{v_{r}^{2}}[1-\beta(r)]$.
        We find:
        
        \begin{eqnarray}\label{eq:cylindrical_dispersion}
                \sigma_{\rho}^{2} &=& \sigma_{r}^{2}\left[1 -\beta(r)\sin^{2}{\theta}\right]\\
                \sigma_{\phi}^{2} &=& \sigma_{r}^{2}\left[1 -\beta(r)\right]\\
                \sigma_{z}^{2}    &=& \sigma_{r}^{2}\left[1 -\beta(r)\cos^{2}{\theta}\right]\label{eq:cylindrical_dispersion3}
        \end{eqnarray}
        These dispersions are related to observable dispersions via the Abel integral \citep{Binney2011-ja}.
        We can then derive the following set of equations \citep{Strigari2007-cc,Van_der_Marel2010-en}:
        \begin{eqnarray}
        \label{eq:los_disp}
            \sigma_{\mathrm{los}}^{2} &=& 
            \frac{2}{I_{\star}(R)}\int_{R}^{\infty}\left(1-\beta\frac{R^2}{r^2}\right)\frac{\nu_{\star}\sigma_{r}^{2} r dr}{\sqrt{r^2 - R^2}}\label{eq:sigma_los} \\
            \label{eq:R_disp}
            \sigma_{\mathrm{pmr}}^{2} &=&
            \frac{2}{I_{\star}(R)}\int_{R}^{\infty}\left(1-\beta+\beta\frac{R^2}{r^2}\right)\frac{\nu_{\star}\sigma_{r}^{2} r dr}{\sqrt{r^2 - R^2}}\\
            \label{eq:losdispersions}
            \sigma_{\mathrm{pmt}}^{2} &=& 
            \frac{2}{I_{\star}(R)}\int_{R}^{\infty}\left(1-\beta\right)\frac{\nu_{\star}\sigma_{r}^{2} r dr}{\sqrt{r^2 - R^2}}.
            \label{eq:t_disp}
        \end{eqnarray}
        Here $\sigma_{\mathrm{los}}$ represents the dispersion along the line of sight, $\sigma_{\mathrm{pmr}}$ represents the dispersion in the plane of the sky in the radial direction from the center of the galaxy, and $\sigma_{\mathrm{pmt}}$ is the dispersion in the plane of the sky perpendicular to $\sigma_{\mathrm{pmr}}$.
        
        Critically, the three observable dispersions (Eq~\ref{eq:sigma_los}-\ref{eq:t_disp}), have a different dependence on $\beta(r)$.
        This will allow us to break the degeneracy between the anisotropy parameter and the density parameters.
        To estimate how well this degeneracy is broken under observations with varying numbers of stars and uncertainties on stellar velocity measurements, we turn to the Fisher Matrix formalism.

\section{Fisher Information}\label{sec:fisher}

    The Fisher Matrix formalism is a way of characterizing the information content within a set of experiments or observations \citep{Fisher1935-rf}.
    This formalism allows us to forecast parameter uncertainties, given a fiducial set of model parameters,
    without running a full computational-intensive Jeans Equation analysis.
    Examples of its recent uses include predicting uncertainties in cosmological parameters \citep{Heavens2009-yh} and forecasting precision in chemical abundance determination from spectroscopic observations of resolved stars \citep{Sandford2020-pd}.

    The Fisher Matrix formalism allows us to quickly forecast an upper bound on the precision with which parameters of interest can be determined by analyzing the likelihood function for a set of observations.
    The only required information is the errors expected from some set of observations and a likelihood function connection observables to the variables of interest.
    Without obtaining the data and without having to do the full modeling we can estimate precisions, determine linear degeneracies, and investigate how these degeneracies change given additional observations/data.
    It can also be used to determine what type of observations and what precision on those observations are needed in order to constrain parameters to a certain level of confidence.
    
    More formally, we assume that the likelihood function is a maximum near its true parameter values $\vec{\theta}_{0}$ and is well estimated by a Gaussian function.
    We can then expand the log-likelihood function around its maximum via a Taylor series:
    \begin{equation}\label{eqn:likelihood_expansion}
        \ln \mathcal{L}=
        \ln \mathcal{L}(\overrightarrow{\theta_{0}})+
        \left.\frac{1}{2} \sum_{i j}\left(\theta_{i}-\theta_{i, 0}\right) \frac{\partial^{2} \ln \mathcal{L}}{\partial \theta_{i} \partial \theta_{j}}\right|_{\theta_{0}}\left(\theta_{j, 0}-\theta_{j}\right)+...
    \end{equation}
    The second term of this expansion characterizes the curvature of the function around its peak.
    The sharper this peak is, the more constraining a set of observations is.
    With that motivation we define the Fisher matrix as the expectation value of the second derivative of the log-likelihood function: 
    \begin{equation}\label{eqn:fisher_definition}
        F_{i j}=
        -\left\langle\frac{\partial^{2} \ln \mathcal{L}}{\partial \theta_{i} \partial \theta_{j}}\right\rangle.
    \end{equation}

    The errors on the parameters $\theta_{i}$ are related to the Fisher matrix by the Cramer-Rao relation~\citep{Rao1945-ep,Cramer1946-np} which states that:

        \begin{equation}\label{eqn:fisher_uncertainties}
            \sigma^{2}_{\theta_{i}} \ge 
            F^{-1}_{ii}.
        \end{equation}
        Equality is obtained when the observables have Gaussian errors, therefore calculating the Fisher matrix in this limit provides us with a lower bound on these uncertainties.

    \subsection{Fisher for Jeans} 
        \label{sub:fisher_for_jeans}
        In the context of spherical Jeans modeling, data is generally fit to some model by maximizing the likelihood function for these observations, which then returns a set of best fitting model parameters with associated uncertainties \citep{Hogg2010-ht}.
        The Fisher formalism, in contrast, allows us to predict the errors on these model parameters by analyzing the likelihood function alone.
        
        We construct the likelihood following \citet{Strigari2018-fz}.
        The probability of observing a stellar velocity $v_i$ assuming a Gaussian measurement uncertainty $\delta v_i$ is given by
        \begin{equation}
            G(v_i,u_i,\delta v_i) = \frac{1}{\sqrt{2 \pi \delta v_{i}^{2}}} \exp \left[-\frac{\left(v_{i}-u_{i}\right)^{2}}{2 \delta v_{i}^{2}}\right].
        \end{equation}
        The distribution for the true velocities $u_i$ is assumed to be Gaussian,
        \begin{equation}
            G\left(u_{i}, \bar{u}, \sigma_{i}\right)=\frac{1}{\sqrt{2 \pi \sigma_{i}^{2}}} \exp \left[-\frac{\left(u_{i}-\bar{u}\right)^{2}}{2 \sigma_{i}^{2}}\right].
        \end{equation}
        While the true likelihood is more complex, the Gaussian model represents a good approximation~\citep{Strigari:2010un}.

        The probability for a measured velocity, given the  model parameters, is then a convolution of the previous two equations:
        \begin{equation}
            \begin{aligned}
                P\left(v_{i} \mid u_{i}, \sigma_{i}\right) &=\int G\left(v_{i}, u_{i}, \delta v_{i}\right) G\left(u_{i}, \bar{u}, \sigma_{i}\right) d u_{i} \\
                &=\frac{1}{\sqrt{2 \pi\left(\sigma_{i}^{2}+\delta v_{i}^{2}\right)}} \exp \left[-\frac{\left(v_{i}-\bar{u}\right)^{2}}{2\left(\sigma_{i}^{2}+\delta v_{i}^{2}\right)}\right].
            \end{aligned}
        \end{equation}
        This equation can be written separately for both radial velocities (los) and the two proper motions components (pmr, pmt).
        We can therefore write a likelihood function for each velocity component $k \in$ (los, pmr, pmt) as:
        \begin{equation}\label{eq:likelihood_individual}
            \mathcal{L}_k = \prod_{i=1}^{n_k} \frac{1}{\sqrt{2 \pi\left(\sigma_{k,i}^{2}+\delta v_{k,i}^{2}\right)}} \exp \left[-\frac{\left(v_{k,i}-\bar{u}\right)^{2}}{2\left(\sigma_{k,i}^{2}+\delta v_{k,i}^{2}\right)}\right].
        \end{equation}
        The full likelihood is a product of the three likelihood functions $\mathcal{L}= \prod \mathcal{L}_k$.
        The Fisher matrix is then:
        \begin{equation}
            F_{ij}
            =\sum_k \sum_{l}\left[\frac{1}{\varepsilon_{k,i}^{2}} \frac{\partial v_{k, l}}{\partial \theta_{i}} \frac{\partial v_{k,l}}{\partial \theta_{j}}+\frac{1}{2} \frac{1}{\varepsilon_{k,i}^{4}} \frac{\partial \sigma_{k,l}^{2}}{\partial \theta_{i}} \frac{\partial \sigma_{k,l}^{2}}{\partial \theta_{j}}\right],
        \end{equation}
        where $\varepsilon_{k,i}^2 =\left(\sigma_{k,i}^{2}+\delta v_{k,i}^{2}\right) $.
        If we assume that there is no net rotation or streaming motion, then  the first moment of the velocity distribution function is zero and this term vanishes.
        We are left with:
        \begin{equation}\label{eq:fisher_matrix}
            F_{ij}
            =\sum_{k} \sum_{l}
            \left[
            \frac{1}{2} \frac{1}{\varepsilon_{k,i}^{4}} \frac{\partial \sigma_{k,l}^{2}}{\partial \theta_{i}} \frac{\partial \sigma_{k,l}^{2}}{\partial \theta_{j}}\right].
        \end{equation} 
        This equation for the Fisher matrix is in the same form as the case for binned observational data as was the case in \citet{Strigari2007-cc}.
        
        The intrinsic velocity dispersion is a function of the model parameters through the projected Jeans Equations.
        We define this set of model parameters as $\vec{\theta}=(\ln \rho_0,\ln r_0,a,b,c,\beta)$.
        We  calculate the expression for our Fisher matrices as a function of the parameters.

        In addition to our base set of parameters, we are interested in the uncertainties and covariances of certain functions of these parameters, for example the log-slope at a given radius or the enclosed mass within a given radius.
        We can use the Fisher matrix formalism to explore such quantities of interest, and examine how our knowledge of them varies as we gather more information.
    
        In particular, consider any function $f$ of our model parameters, defined as $f(\vec{\theta})$.
        We can calculate the covariance matrix, $C^f$, as
        \begin{equation}\label{eq:errors}
            C^{\mathrm{f}}=\mathrm{J} C^{\vec{\theta}} \mathrm{J}^{\top}= \mathrm{J} F^{-1} \mathrm{J}^{\top},
        \end{equation}
        where J is the Jacobian matrix given by $J_{ij}=\frac{\partial f_i}{\partial \theta_j}$ and $C^{\vec{\theta}}=F^{-1}$ is the inverse of the Fisher matrix given by Equation~\ref{eq:fisher_matrix}.
        We then determine the pair-wise confidence regions by calculating a $\chi^2$ surface:
    
        \begin{equation}\label{eq:chi2}
            \chi^2 = (f-f_{\mathrm{model}})(C^f)^{-1}(f-f_{\mathrm{model}})^{T}
        \end{equation}
        where $\Delta \chi^2 = 2.3,6.17,11.8$  corresponds to $1\sigma,\,2\sigma$ and $3\sigma$ respectively.
        Equation~\ref{eq:chi2} may be generally used to compute errors on our base set of parameters, or on parameters derived from them as we discuss in Section~\ref{ssub:mass}~-~\ref{ssub:log_slope}.
    
    \subsection{Fisher Computation}\label{sub:fisher_computation}
        We now have  the information needed to calculate the Fisher matrices.
        To calculate the derivatives with respect to model parameters, we numerically differentiate the projected Jeans Equations.
        For the example case of the line-of-sight component, this takes the form: 
        \begin{equation}
            \label{eq:num_derivative}
            \frac{\partial \sigma^2_{\mathrm{los}}}{\partial \theta_{i}}
            =\frac{\sigma^2_{\mathrm{los}}\left(R, \theta+\Delta \theta_{i}\right)- \sigma^2_{\mathrm{los}}\left(R, \theta-\Delta \theta_{i}\right)}{2 \Delta \theta_{i}}
        \end{equation} 

        There are similar expressions for the pmr and pmt components.
        Our analysis does not take into account the possibility of systematic errors in the observed measurements which can in principle be a large source of uncertainty~\citep{Read2021-du}.
        
    \subsection{Priors}
        Including all three velocity components, the final expression for our Fisher matrix can be written as:
        \begin{equation}
            F = F_{\mathrm{los}}+ F_{\mathrm{pmr}} + F_{\mathrm{pmt}} + F_{\mathrm{prior}},
        \end{equation}
        where the last term, $F_{\mathrm{prior}}$ represents the priors on our parameters.
        We take our parameters to be $\vec{\theta}=\{\ln\rho_{0},\ln{r_{0}},a,b,c,\beta\}$ so that each parameter is dimensionless.
        For simplicity, throughout this work we set Gaussian priors of unit variance on every parameter, i.e., $F_{\mathrm{prior}}$ is the identity matrix.
        While flat priors \citep[e.g.,][]{Acquaviva2012-xu,Read2017-nl,Read2018-bv,Read2021-du,Hayashi2020-cc,Chang2021-bu} are used in practice, Gaussian priors allow us to do this calculation analytically.
        Since priors can bias the results of an MCMC analysis we analyze choice of priors in Section~\ref{ssub:priors} and show that our final results do not depend strongly on the choice of priors when the kinematic samples are larger than 100 stars.
        
        With this formalism defined, we can forecast uncertainties on parameters.
        Since the results of this analysis will vary depending on the choice of fiducial parameters, priors, velocity precision, and location of stars, we will make the code accessible at \href{https://github.com/dmForecast/dmForecast}{dmForecast}\footnote{\url{https://github.com/dmForecast/dmForecast}}.
        In the next sections we state some of the main results that can be done with this tool.

\section{Analysis}\label{sec:analysis}
    The Fisher matrices give us the uncertainties on our base set of model parameters: $\vec{\theta}=\{\ln\rho_{0},\ln{r_{0}},a,b,c,\beta\}$.
    As discussed above, the Fisher matrix does not depend on the values of the data, but it does depend on the model parameters.
    This implies that the errors depend on the specifics of a particular system.
    For our fiducial set of model parameters, we choose to roughly reproduce the observed velocity dispersion of the Draco dSph \citep{Walker2009-fv}, which has a nearly constant line-of-sight velocity dispersion of $\sim 10$\,\kms ($\vec{\theta}_{\mathrm{cusp}}=\{16.8, 0.69, 1, 1,3,0\}$, $\vec{\theta}_{\mathrm{core}}=\{18.8,0.4, 0, 1.5, 3, 0\}$).

    An additional critical piece of information is the observed or expected measurement errors on the stellar line-of-sight velocities and proper motions.
    For Milky Way dSphs, current observed radial velocity samples used to determine kinematics are observed at medium spectral resolution (R $\sim 5000 -- 20,000$) with typical measurement errors between $\sim 2 - 5$\,\kms \citep[e.g.,][]{simon2007, Walker2009-fv, Battaglia2013-we}.
    Looking to the future, we focus on the next generation of medium resolution spectrographs on 30-meter class telescopes which will provide large samples of stars with similar velocity errors (2\,\kms), but for fainter stars over a wider field of view \citep{wolff2019}.
    Current observational errors on individual proper motions are large compared to radial velocities, with the best current internal proper motion errors of order 20\,\kms (0.05 mas/year) \citep{Massari2020-ue}.
    The next generation of 30-meter class telescopes with adaptive optics-enabled imaging are expected to produce improved transverse velocity errors of $\sim 5$\,\kms \citep{Kallivayalil2015,simon2019-ba}.
    While we present results using on these measurement errors, our method is flexible and can easily accommodate a different set of observational uncertainties.

    \subsection{Enclosed Mass}\label{ssub:mass}
        \begin{figure*}[htp!]
            \centering
            \includegraphics[width=.99\linewidth]{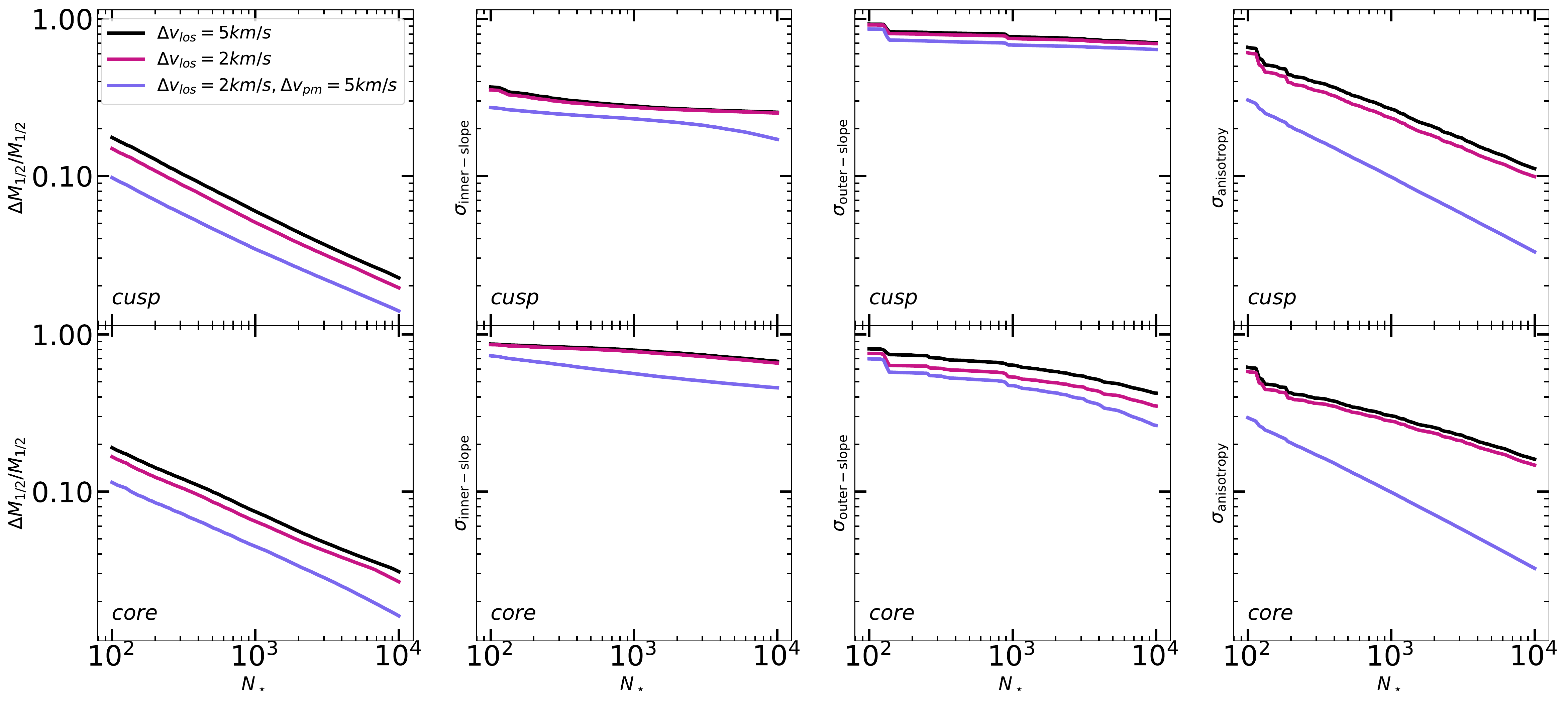}
            \caption{
                The uncertainty in parameters as function of number of stars ($N_\star$) with velocity data for a Draco-like dSph.
                The black and red lines are for samples of radial velocities with velocity error 5\,\kms and 2\,\kms.
                The blue lines are for equal number of radial and proper motion measurements.
                The upper panels show results for a fiducial cusp profile, while bottom panels are a core profile.
                }
        \label{fig:uncertainties}
        \end{figure*}
    
        The enclosed mass is one of the few quantities that can be well constrained with only radial velocities~\citep{Strigari2007-cc,Walker2009-fv,Wolf2010-xe}.
        We first use our framework to explore the improvements on the enclosed mass uncertainties from increasing the number of stars alone.
        In Figure~\ref{fig:uncertainties} we draw stars from a Plummer model with a scale radius of $r_p = r_{1/2} = 250$ pc.
        We plot the resulting parameter uncertainties for 3 different samples:  stars with radial velocity errors of 5\,\kms\ (black), radial velocity errors of 2\,\kms\ (red), and combining radial velocities (2\,\kms\ errors) and proper motion measurements with 5\,\kms errors (blue).
        
        In the first column of Figure~\ref{fig:uncertainties}, the enclosed mass at $r_{1/2}$ is already well constrained with $\sim 100$ radial velocity measurements alone.
        The uncertainties decrease steadily with increased sample size.
        Going from $5~\kms$ to $2~\kms$ precision on radial velocities does not yield significantly better results, compared to the addition of proper motions.
        From $\sim 10^3$ stars to $\sim 10^4$ stars, we see $\sim 60\%$ improvement to the mass uncertainty.
        At $\sim 10^4$ stars we are able to measure the mass with a few percent error with RVs.
        Adding a similar number of proper motion stars improves the predicted errors on enclosed mass by roughly $30\%$.  Uncertainties on the enclosed mass remain the same for a cusp or cored profile (top and bottom panel of Figure~\ref{fig:uncertainties}, respectively).

\begin{figure*}[htp]
            \includegraphics[width=.99\linewidth]{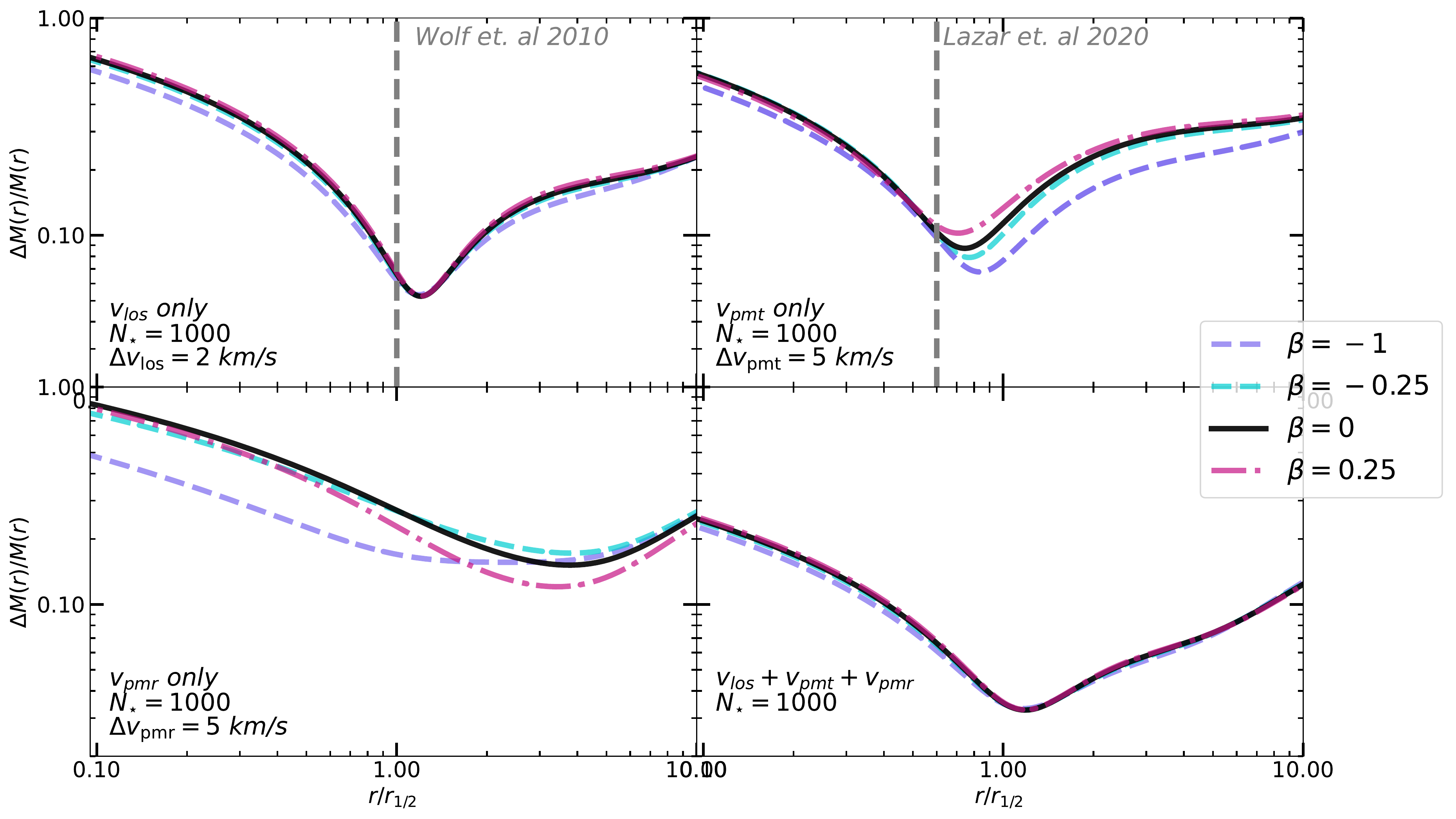}
            \caption{Error on the enclosed mass as a function of radius.
            Clockwise from top left to bottom right, the results for 1000 RVs velocities with $\delta v= 2$ \kms, 1000 pmt or 1000 pmr velocities with $\delta v = 5$ \kms, and all three velocity components combined.
            The vertical grey lines in the top left and right panels indicate the radius where the Wolf-mass estimator and \citet{Lazar2020-jr} estimator are defined.Colored lines indicate models with different values of anisotropy ($\beta$).
            From the minimum values in each panel, we conclude that the enclosed mass can be measured within 5\% with RVs velocities alone, and 3\% if we include both RVs and proper motion.}
            \label{fig:mass_error}
        \end{figure*}

        We next fix the number of stars to 1000 and plot the fractional error on the enclosed mass as a function of radius for our fiducial NFW model (Figure~\ref{fig:mass_error}).
        We show each of the observable velocity components separately.
        While the pmt and pmr velocities will always be measured jointly, it is informative to investigate their properties separately.
        As in Figure~\ref{fig:uncertainties}, we draw stars from a Plummer model with a scale radius of $r_p = r_{1/2} = 250$ pc.
        
        The top left panel of Figure~\ref{fig:mass_error} shows the uncertainty in the enclosed mass as a function radius for a sample of stars with radial velocities only.
        We find that the enclosed mass is well constrained (within $5\%$) with radial velocities alone if measured around $r\sim r_{1/2}$.
        This radius is very close to that defined in the popular ``Wolf mass" estimator \citep[eq. 1-2]{Wolf2010-xe}, as well as the mass estimator defined in~\citep[eq. 11]{Walker2009-fv} and~\citep[eq. 12]{Errani2018-rf}.
        The Wolf mass radius was analytically determined to be the radius which the anisotropy has a minimal effect on the mass.
        We find that this radius minimizes the total errors on {\it all parameters} of the fit, not just anisotropy.
        This is consistent with findings from simulations \citep{Gonzalez-Samaniego2017-ae,Campbell2017-nx,Applebaum2021-lf}.
        The minimum radius varies slightly depending on the choice of parameters, but it is always near $r_{1/2}$.
        We find similar behavior when considering transverse pmt velocities.
        The fractional error is minimized at or slightly less than $r_{1/2}$.
        This information can be used to independently measure the mass within $10\%$ with a velocity precision of $\delta v_{\mathrm{pmt}}= 5$ \kms.
        We do find a stronger dependence on the stellar anisotropy as compared to RVs alone, although in every case, there is still a pronounced minimum.
        Similar to the Wolf mass estimator, this minimum radius agrees with the expression derived analytically by \citet{Lazar2020-jr}.
    
        It is interesting to note that the pmr velocities display different behavior than the other components.
        In particular, the error does not have a strong minimum as with the previous two components.
        In this case, the shape of the error on the enclosed mass varies significantly as $\beta$ is varied.~\cite{Lazar2020-jr} came to a similar conclusion and for this reason did not consider a mass estimator formula for this component.
        
        In the bottom right panel of Figure~\ref{fig:mass_error}, we show the results when combining all three components of the velocities.
        As expected the minimum error is achieved at $r\sim r_{1/2}$; with a sample of 1000 stars, at a precision of $\delta v_{\mathrm{los}} = 2$ \kms and $\delta v_{\mathrm{pmt}} = 5$ \kms, it will be possible to measure the enclosed mass to within $3\%$.
        Uncertainties on the enclosed mass are critical in designing dark matter annihilation experiments in which constraints depend on the dark matter density enclosed with some radius.
        
    \subsection{Dependence on Sample Size}\label{ssub:sample_size}
        In Figure~\ref{fig:uncertainties} we show how increasing the number of observed stars affects the uncertainties on the model parameters of interest.
        In the previous section we discussed the left columns of this figure, noting the the uncertainties on the enclosed mass decrease steadily with increased sample sizes.
        We find a similar trend in the stellar anisotropy, $\beta$ (right columns, Figure~\ref{fig:uncertainties}).
        However, with radial velocities alone the uncertainty on $\beta$ decreases only marginally as a function of $N_\star$, never falling below $\sigma_\beta =0.1$.
        
        Our model predicts a drastic change in the uncertainty in $\beta$ when adding proper motions to the analysis.
        This result is driven by the three equations presented in Section~\ref{ssub:solution_jeans_equation} (Equations~\ref{eq:cylindrical_dispersion}-\ref{eq:cylindrical_dispersion3}), in which the defining difference between the observable velocity dispersions is their differing dependencies on the stellar anisotropy.
        This makes them incredibly useful in constraining this parameter as well as removing the degeneracy that is usually present when only RVs are available.

        Though the focus on our analysis is the central slope of the dark matter profile, it is useful to note how our formalism may be used to constrain the slope in the transition regime, $b$, and the asymptotic outer slope, $c$.
        This is an interesting regime because it determines the extent to which the outer region of the dark matter has been stripped from the dSph by the tidal field of the Milky Way.
        In the context of our Fisher matrix formalism, it can also provide insight into how degenerate these parameters are in determining the central slope $a$.
        In the case of the inner-slope and outer-slope, we find that these variables are challenging to constrain.
        The uncertainty in these two parameters decreases at a significantly smaller rate than that of the mass and anisotropy.
        
        This ability to constrain $b$ and $c$ is driven by the numbers of stars at large radii.
        For our Plummer model, when drawing a total of $\sim 10^4$ stars, there are $\sim 500$ stars at radius $> 1$ kpc, and only $\sim 100$ beyond 2 kpc.
        In Figure~\ref{fig:uncertainties}, we see that the uncertainty on $c$ does not change significantly from the priors we set, not falling below $\sigma_c \sim 0.65$, regardless of how many stars we consider.

    \subsection{Dependence on Observational Errors}\label{ssub:errors}
        A natural assumption in planning upcoming surveys is that improved measurement errors on individual stars will contribute to improved constraints on the derived dark matter density profiles.\
        Our framework explicitly predicts how such improvements will ultimately tighten our constraints.
        In Figure~\ref{fig:uncertainties}, we find only a slight improvement on the model parameter uncertainties as we decrease the uncertainty on radial velocities, $\delta v_{los}$, from $5$ \kms (black) to $2$ \kms (red).
        Decreasing $\delta v$  further has a negligible effect, with similar results for the transverse measurements.
        We can understand this by analyzing the likelihood function (Equation~\ref{eq:likelihood_individual}).
        Assuming that all three observable velocity dispersions are $\sim 10$ \kms, the intrinsic dispersion term completely dominates over the velocity measurement error.
        This is explicitly shown  in the likelihood function where the velocity measurement error always appears as $(\sigma^2 + \delta v_i^2)$.
        Thus, further decreasing the errors, or even assuming no errors, does not impact parameter uncertainties.
        However, systems with lower velocity dispersions, for example Segue 1 with a velocity dispersion $\sim 3-4$ \kms \citep{Simon2011-dk}, will be more strongly affected by the velocity uncertainties that are achievable with present and future measurements.
        
        In principle one could add an additional error term in the likelihood function so that the total error budget consists of the intrinsic dispersion, observational (random) uncertainties on the velocities, and an additional term to account for systematic errors.   

    \subsection{Dependence on Priors}\label{ssub:priors}
        \begin{figure}[t] 
            \centering
            \includegraphics[width=.95\linewidth]{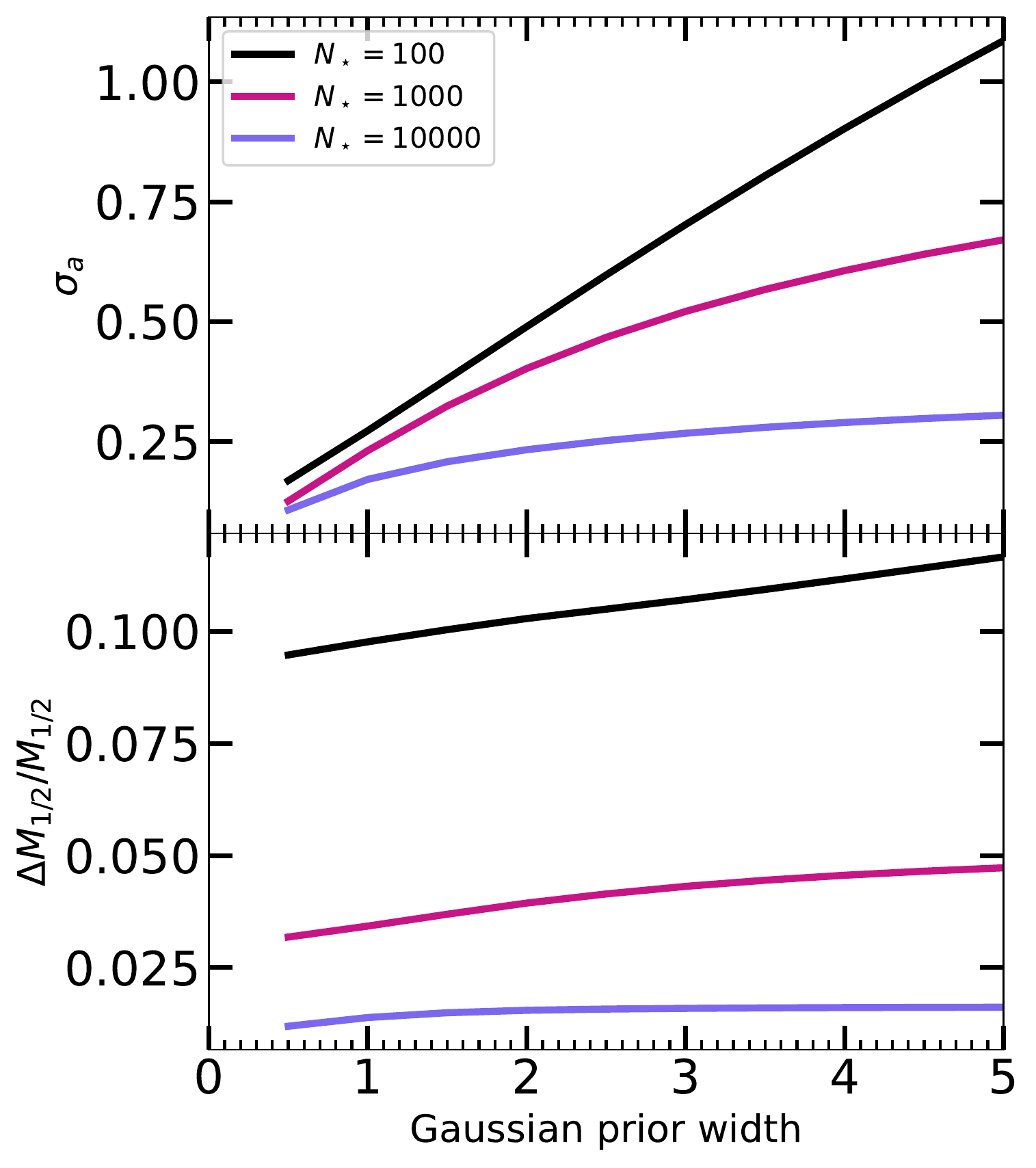}
            \caption{We examine how our choice of priors affects the predicted uncertainties of our analysis.
            In each panel, we plot the predicted uncertainties against the width of our Gaussian priors.
            ({\it Top panel:}) The uncertainty on the inner-slope is significantly affected by the width of the Gaussian prior when $N_\star<100$.
            However, as $N_\star$ grows towards 10,000 RVs + PMs the priors no longer affect the results significantly.
            ({\it Bottom panel:}) The uncertainties on the enclosed mass are less dependent on the choice of priors for all sample sizes plotted.}
            \label{fig:priors}
        \end{figure}
        The choice of priors is known to influence the parameter estimation in the analysis of dSph density profiles~\citep{Martinez2009-dm,Martinez2015-bc}.
        When the sample size is small, the constraints provided by the data are not strong enough to overcome the arbitrary choice of the assumed priors.
        For our baseline analysis throughout this paper, we utilize Gaussian priors with width $\sigma_{\mathrm{Gauss}} = 1$ (recall we scaled our parameters to be dimensionless).
        We examine the impact of this choice in Figure~\ref{fig:priors}.
        In the top panel of this figure, we vary the width of the prior and compare the results on the uncertainty of the inner slope.
        We find that for small $N_\star$ ($< 100$ stars), the priors play a significant role in the resulting uncertainty.
        As $N_\star$ increases, there is a reduced impact of the priors.
        The uncertainty curve for $N_\star=1000$ is shallower than for $N_\star=100$, and once we get to $N_\star =10000$ the curve is almost flat.
        For $N_\star=100$, $\sigma_a$, the inferred uncertainty increases by a factor of 5  (from $\sigma_a = 0.16$ to $1.08$) as the prior width is varied from $0.5$ to $5$ (Figure~\ref{fig:priors}, black curve), but $N_\star = 10,000$ (blue curve), this increase is minimal.
        
        In contrast, when we consider the uncertainty on the integrated mass, there is a minimal impact from the assumed priors.
        With only $\sim 100$ stars the uncertainty on the integrated mass weakly depends on the priors (bottom panel, Figure~\ref{fig:priors}).
        This reinforces our conclusion in Section~\ref{ssub:mass} that the error on the integrated mass from a spherical Jeans analysis is robust.

        In summary, for all of our base set of parameters, we find that for small samples of stars the results are strongly dependent on priors.
        Our choice of priors leads to an optimistic result at $N_\star \sim 100$.
        The impact of this choice is generally mitigated for sample sizes of 1000 stars or greater.
        Regardless, this analysis has the ability to show how results will depend on changing all of these parameters.

\section{Implications For the Core/Cusp Problem}\label{sec:comparisons}
    \begin{figure*}[htp]
        \includegraphics[width=.99\linewidth]{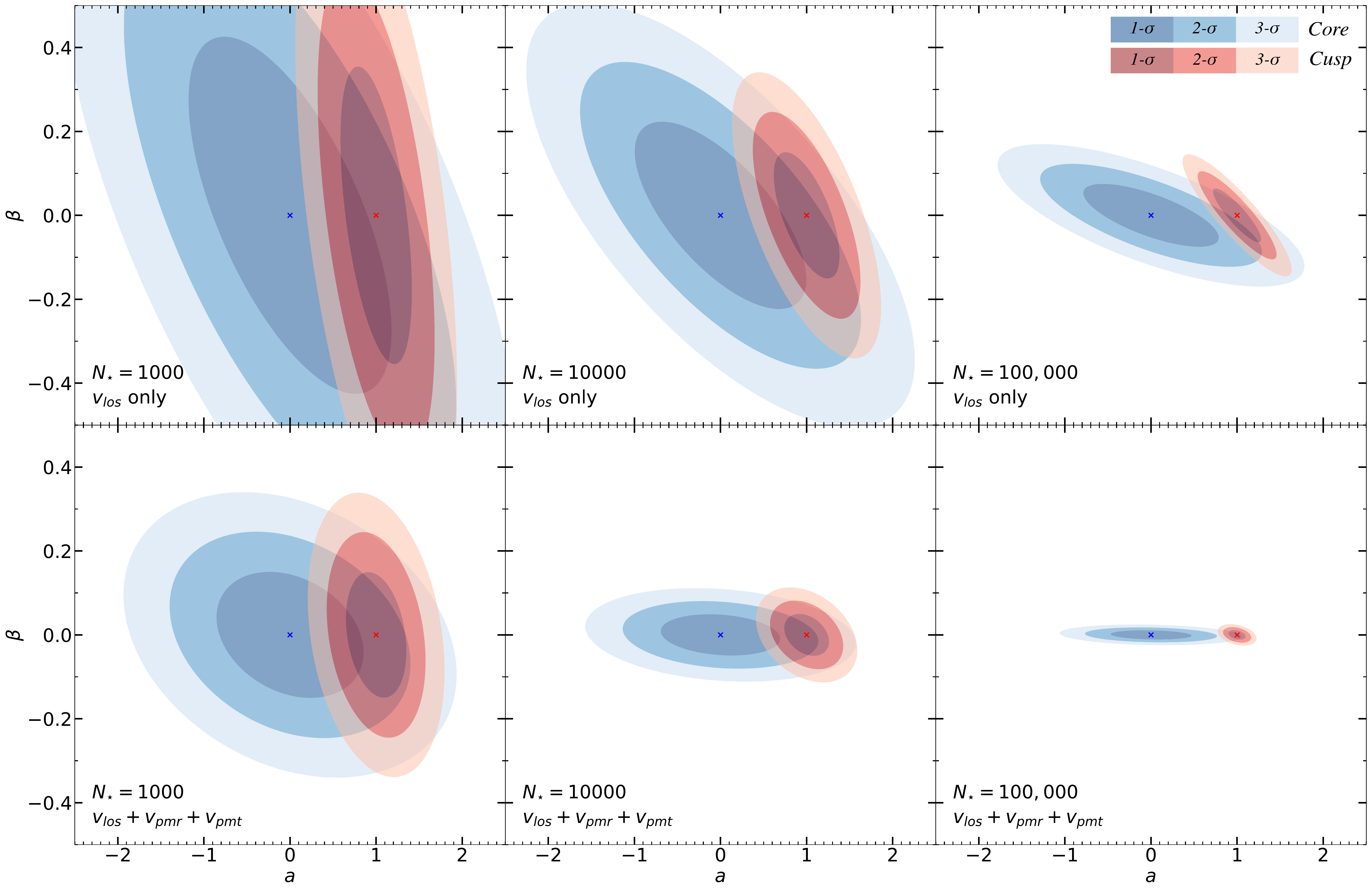}
        \caption
        {
        What observations are required to differentiate between a cored and a cusped density model?
        In each panel, $1\sigma$, $2\sigma$, and $3\sigma$ contours are plotted as a function of the inner slope (a) and anisotropy ($\beta$).
        Blue contours are center at $(a,\beta)=(0,0)$ for our core model and red is centered at $(a,\beta)=(1,0)$ for our cusp model.
        The top rows show increasing samples of radial velocities only ($\delta_v = 2\,$\kms).
        The bottom row shows the case where we have RVs and proper motions ($\delta_{pm} = 5\,$\kms).
        It is only in the bottom rightmost panel (100,000 stars with all 3 velocity components) that the regions corresponding to two sigma uncertainties no longer overlap.
        } 
        \label{fig:contours}
    \end{figure*}

    We now use our formalism to answer the question: What types of observations are necessary to constrain the dark matter central densities and, more specifically, can we distinguish between cusps and cores?

    \subsection{Inner Density Slope}\label{sub:inner_slope} 
        In the second column of Figure~\ref{fig:uncertainties}, we showed the constraint on  the inner slope as a function of the number of observed stars.
        Our model predicts that even with a large sample of stars, RVs alone will not be enough to constrain the inner slope, as expected.
        In particular, when increasing the number of stars from $\sim 10^2$ to $10^3$, the uncertainty on the inner slope saturates at $\sigma_a \sim 0.3$ for the our cusp model.
        This is in general agreement with previous studies utilizing RVs alone~\citep{Lokas2002-lk,Hayashi2020-cc}.
        Also in agreement with previous studies~\citep[e.g.,][]{Strigari2007-cc,Read2021-du}, the addition of proper motions with errors of $\sim 5$\,\kms significantly increases our ability to constrain the inner slope; for 1000 stars, the uncertainty drops from $\sigma_a\sim0.3$ to $\sigma_a \sim 0.2$, which for our fiducial cusp corresponds to a measurement of this parameter to $\sim 20\%$.
        As we approach $10^4$ stars for our cusp with RVs + PMs, the uncertainty decreases only slightly to $\sim 17\%$.
        As stated in Section~\ref{ssub:errors} we have diminishing returns from increasing the precision our velocity measurements with respect to our intrinsic velocity dispersion.
        At $\sim 1000$ stars, improving from $5$\,\kms to $2$\,\kms errors only produces a change on the order of $10^{-2}$ for $\sigma_a$.
        This is shown for RV measurements in Figure~\ref{fig:uncertainties} and is similar for proper motions errors.

        Our core model is significantly less well constrained in terms of the inner slope (second column, bottom panel Figure~\ref{fig:uncertainties}).
        The error on $\sigma_a$ is consistently a nearly a factor of two larger as compared to our cusp model.
        Since we are using the same stellar profile for both models, we expect fewer stars to be in the inner region of our core model.
        However, forcing equal number of stars within the scale radius of both models did not make the uncertainties comparable.
        We found that as we decrease the scale radius (see Figure~\ref{fig:densityvlogslope}), and therefore increased the scale density to maintain a roughly similar velocity dispersion, our ability to constrain the inner-slope drastically decreases.
        When the scale radii are roughly the same, both models are about equally constrained~\citep{Strigari2007-cc}.

    \subsection{Differentiating between models}\label{ssub:our_method}
        We now move on to determining specifically how well we can distinguish between our fiducial core and cusp models.
        We progressively increase the number of stars, and consider one scenario in which we can only measure RVs and another where we can measure RVs and proper motions.
        We then determine at which point the cusp and core models become distinguishable, i.e, when the contours describing the allowed regions do not significantly overlap at the $2-\sigma$ level.
        The results of this analysis is shown in Figure~\ref{fig:contours}.

        From left to right, we consider the cases where we have data for $10^3, 10^4$ and an optimistic $10^5$ stars.
        The top row of plots shows the confidence regions corresponding to $1, 2, 3\,\sigma$ for both our fiducial cusp and core when only RVs are available with $\delta_v = 2$\kms.
        With $10^3$ stars, the parameter spaces have considerable overlap.
        As we increase to $10^4$ and $10^5$ stars, the allowed regions of parameter space shrink and begin to separate.
        However, even $10^5$ RV stars are not sufficient to distinguish the different models at $2\,\sigma$ (top right panel, Figure~\ref{fig:contours}).

        The situation improves with the addition of proper motions.
        Our model predicts that with 10,000 stars, the $1-\sigma$ confidence intervals are well separated, though the $2-\sigma$ intervals still overlap.
        It is not until we reach $\sim 100,000$ stars that the $2-\sigma$ confidence regions are well separated.
        The addition of proper motion (with $\delta_{pm} = 5$\kms) shrinks the confidence region in the $\beta$ direction significantly as we would expect from Figure~\ref{fig:uncertainties}.
        The inner-slope also shrinks significantly for our cusp profile.
        However, our core model is never well constrained despite removing the degeneracy with the anisotropy.
        Note that the inner-slope is also degenerate with the scale radius and density, increasing the difficulty in constraining this parameter~\citep[][]{Chang2021-bu}.
            
    \subsection{The log-slope}\label{ssub:log_slope}
    
        \begin{figure*}[htp]
            \includegraphics[width=.99\linewidth]{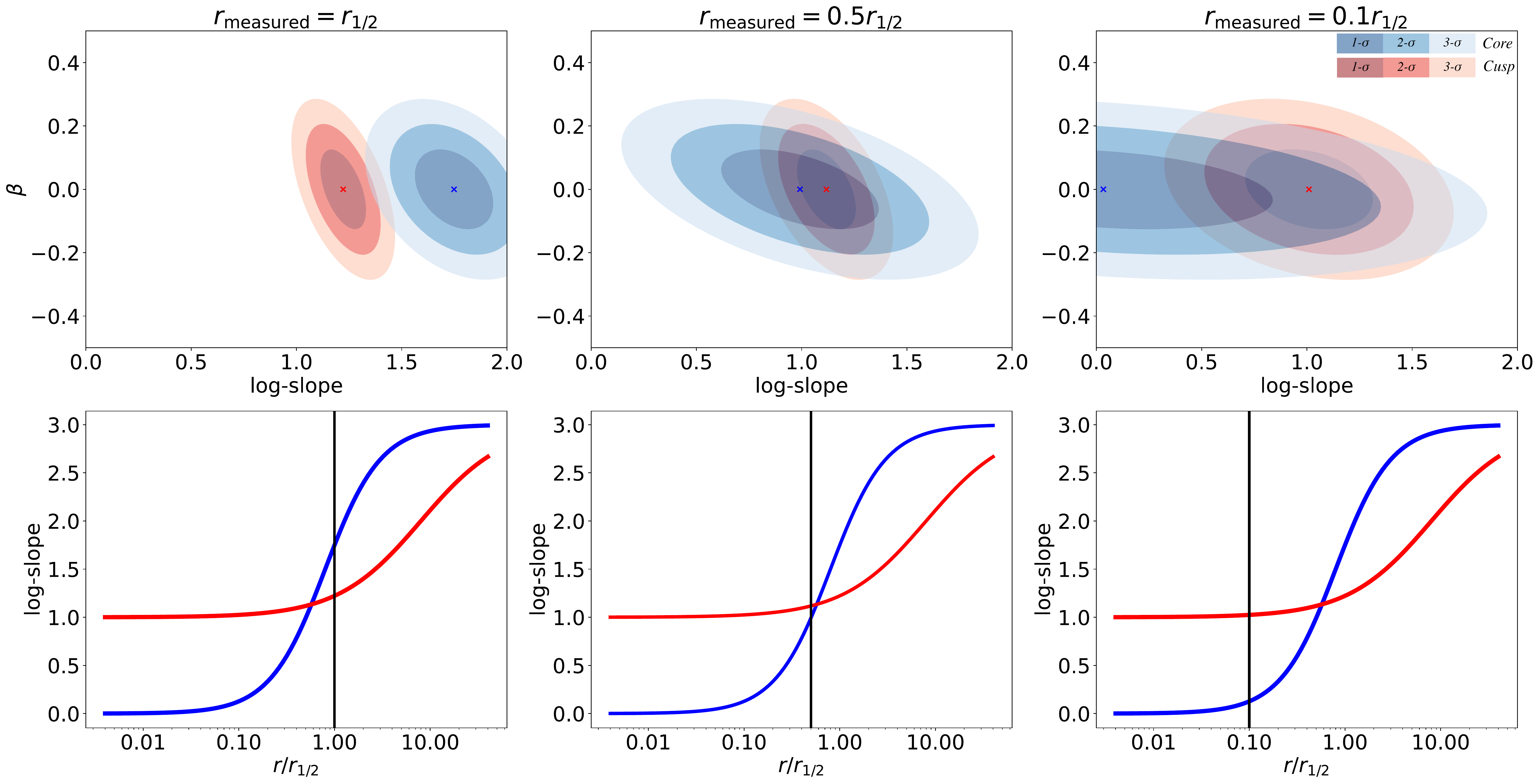}
            \caption{
                Confidence intervals for the log-slope at various radii for 1,000 stars, including both RVs and PMs.
                When measured at $r_{1/2}$ (vertical black line) this parameter is very well constrained (left panel).
                However, for the core profile at $r_{1/2}$ we are not probing the asymptotic limit of the log-slope.
                This is because the scale radius of our core model is the same order of magnitude as $r_{1/2}$.
                Measuring the log-slope is much more effective at a much smaller radius ($r < 0.1 r_{1/2}$).
                At this radius the log-slope becomes unconstrained as is the inner-slope at this number of stars.}
            \label{fig:density_log_slope}
        \end{figure*}
        
        The log-slope of the dark matter density profile ($\gamma(r)= \frac{\partial \ln\rho}{\partial\ln r}$) is often used as an analog for the inner-slope~\citep{Strigari2007-cc,An2009-zo,Van_der_Marel2010-en,Read2018-bv}.
        Similar to~\cite{Strigari2007-cc}, we find that the uncertainty on $\gamma$ is error to have a minimized near the effective radius $r_{1/2}$, which is roughly equivalent to Strigari et al.'s~$r\sim2\,r_{\mathrm{king}}$.
        Differences in this analysis and others differ mainly due to the choice of priors and stellar distribution, in particular King vs.~Plummer stellar density profile.

        In Figure~\ref{fig:density_log_slope}, we explore constraining the log-slope at three different radii.
        Most papers report values for this parameter at $r_{1/2}$ (left panel).
        While the log-slope is well constrained at this radius, it is not always informative.
        The log-slope for our cusp model at $r_{1/2}$ (red line, left panel, Figure~\ref{fig:density_log_slope}) is well within the scale radius and is therefore probing the inner-slope.
        In contrast, our core model (blue lines, left panel) is not probing the inner regions of the halo since our core model has $r_{1/2} = r_0$.
        This could be misleading as the measured value in this case is close to that of a cusp.
        As expected, if we measure $\gamma$ at smaller radii (middle and right panels), the confidence regions approaches that of the inner slope, becoming more difficult to constrained with currently available data.

    \subsection{Previous Work}

        In Section~\ref{ssub:our_method} we discussed what type of observations would be necessary to differentiate between a cusped and cored system.
        We used our Fisher formalism to explore when the posterior parameter space is separated enough such that the 2D confidence regions in the $\beta-a$ dimension are no longer overlapping.
        There has been considerable amount of work addressing this problem using mock observations, we highlight several recent papers below, then compare the strengths of mock observations versus our Fisher analysis approach.
        
        \cite{Chang2021-bu} used mock observations to examine the posterior distribution of inner-slopes.
        They found that even with 10,000 RVs, there was a non-negligible probably that $a=0$ when they tried to constrain the inner-slope for their cusped mock data.
        They did find that the core was better constrained when looking at the posterior distribution of $a$'s.
        However, they then considered the Bayes Factor (BF) to determine whether a cusp or cored would be favored.
        Using method, they found that in both the core and cusp mock data the true value was favored over the alternative.
        However, these predictions may be optimistic considering that they used a simplified model ($\{\rho_0,r_0,a\}$) in which the stellar anisotropy, $\beta$, was not a parameter and so they did not have to account for the $\rho-\beta$ degeneracy.
        For their cusp mock data, their errors for 10,000 RVs are comparable to our errors when using 10,000 RVs + PMs, as expected considering that having proper motions is almost equivalent to eliminating $\beta$ as a parameter while simultaneously providing additional information to constrain the other parameters -- differences in the actual value are then due to differences in parameters considered and priors.

        \citet{Read2018-bv} considered mock profiles of a Draco-like galaxy under various conditions: a pristine NFW cusp profile, one where the cusp was transformed into a core due to bursty star formation, and a self interacting dark matter model.
        Considering various scaling relations with respect to the outer regions of the halo, they found that a single measurement measurement of the density magnitude at $\sim 0.75 r_{1/2}$ could be used to rule out a core and other DM models.
        These authors utilize RVs with virial shape parameters (VSPs), which provides a separate method to break the degeneracy with the velocity anisotropy.
        However, using fourth order moments restricts the nature of the underlying distribution function more so than when using second order moments for both RVs and proper motions~\citep{1990AJ.....99.1548M}.
        \citet{Read2018-bv} also considered considered a more restricted set of cores than this work which explains the result that a single measurement of the density is enough to rule out a core.

        \cite{Read2021-du} applied four different dynamical mass modeling methods to mock data.
        Similar to our analysis, they found that breaking the degeneracy between the shape of the density profile and the stellar anisotropy requires additional information beyond radial velocities.
        That work found that for isotropic or tangentially anisotropic systems, all methods behaved similar well, but radially anisotropic systems proved more difficult and were better handle by methods that used information about the shape distribution function which Jeans modeling does not.
        This limitation can also be probed with the Fisher approach, but we leave this for future work.
        
        Mock observations studies are highly complementary to our Fisher analysis.
        Our Fisher analysis is limited by the choice of Gaussian priors on our parameters where as a full analysis on mocks is free to choose any sort of priors, which can in turn change the results of the analysis.
        Although the Fisher formalism can capture various aspects of analysis,  mock observations are better tool for in-depth analysis.
        However, mock observations become expensive to test various combinations of parameters and observational errors as samples become large.
        Thus, the Fisher analysis presented here may prove a useful first tool, before committing to a more careful analysis using mocks. 

\section{Validation}\label{sec:validattion}

    To validate our Fisher predictions, we make use of the mock data suite from the\dataset[Gaia Challege wiki]{http://astrowiki.ph.surrey.ac.uk/dokuwiki/doku.php?id=tests:sphtri:spherical}\footnote{http://astrowiki.ph.surrey.ac.uk/dokuwiki/doku.php}.
    We will compare our predictions on parameter uncertainties to those directly determined from mock data.
    We use mock data for a spherical, single component, isotropic dark matter halo with an NFW dark matter density profile and a Plummer stellar density.
    These variables produce a velocity dispersion similar to that of the Draco dSph ($10-15$\,\kms).
    
    We built a Jeans modeling fitting code using Dynesty~\citep{Speagle2020-wm} to sample the likelihood defined in Equation~\ref{eq:likelihood_individual}.
    We limit the number of parameters to the inner slope, scale radius and scale density of the dark matter halo.
    We place the same Gaussian priors as in our Fisher analysis, but cut these off to exclude un-physical parameters.
    
\begin{figure}[htp]
    \centering
    \includegraphics[width=.99\linewidth]{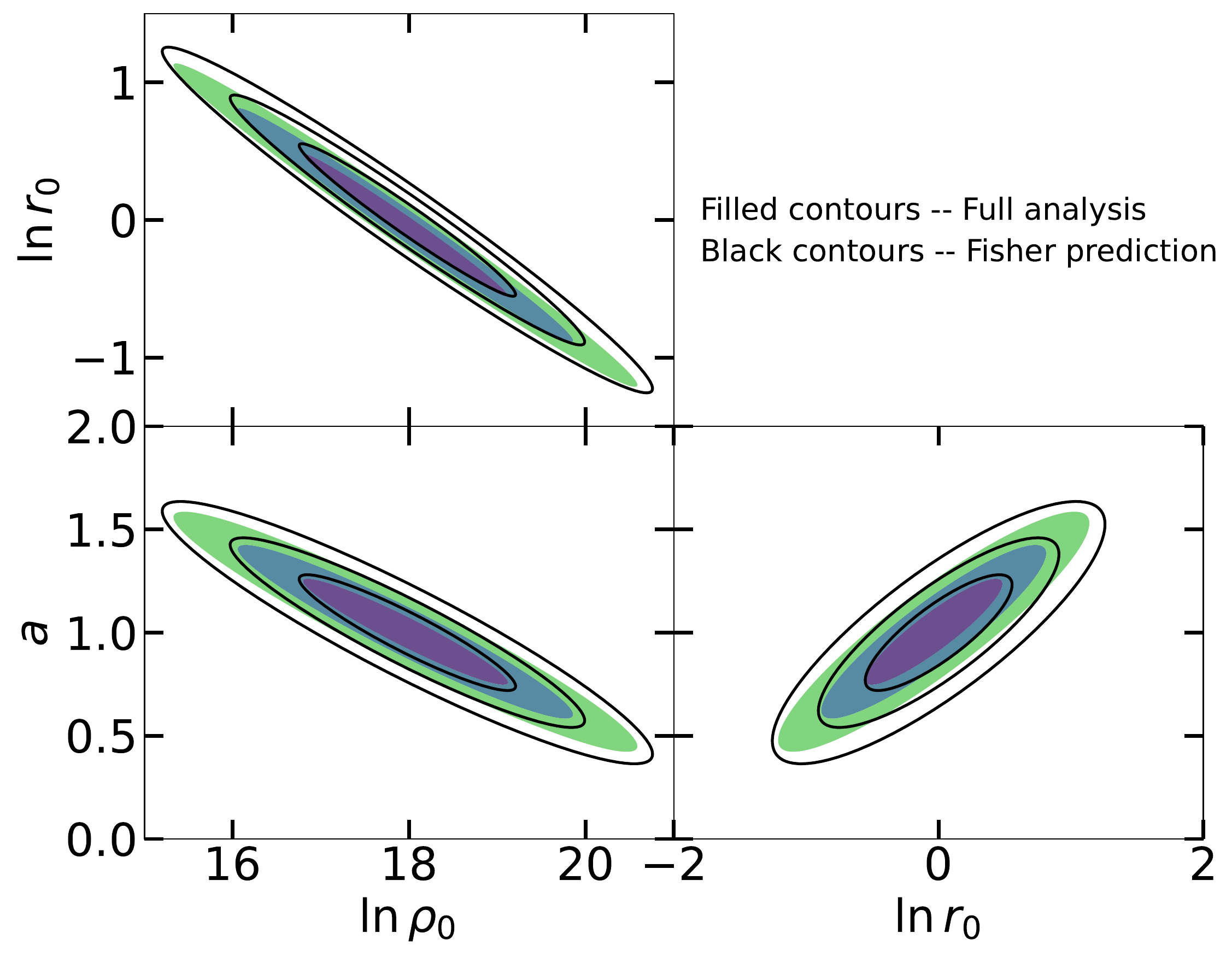}
    \caption{A comparison of our Fisher predictions (black contours) to a full Jeans analysis run on a mock kinematic data set (filled contours) for 1000 stars with $\delta v = 0$.
    Confidence regions are shown for the  scale density, scale radius and inner slope.
    The Fisher analysis accurately predicts the magnitude of the uncertainties as well as the orientation of the confidence regions, validating our methodology.}
    \label{fig:fisherVdynesty}
    \end{figure}

    Figure~\ref{fig:fisherVdynesty} shows the results of this comparison.
    The Fisher Matrix formalism agrees with the errors obtained on model parameters from a full Jean modeling of the likelihood, providing confidence that we can accurately estimate model parameter uncertainties without the need for simulated data.
    The Fisher estimates (black contours) for the 2D confidence regions agree well with those obtained from the analysis of the mock data set (color contours).
    The Fisher formalism accurately predicts the size, shape and orientation of these confidence regions.
    We repeated this analysis on a core model and found similar agreement.
    The small differences at the 2 and 3-$\sigma$ boundaries are due to the fact that we used truncated Gaussian priors in our analysis.
    Various implementations of spherical Jeans modeling are available~\citep{Mamon2013-bi,Cappellari2015-jt,Read2017-nl} and while those methods may be more sophisticated, we expect the predictions from our Fisher approach should be insensitive to the implementation of Jeans modeling used.

    In most implementations, Fisher analysis does not capture non-Gaussianity in the posterior distributions due to non-linear degeneracies.
    Previous studies have shown that degeneracies found in Jeans analysis are approximately linear for $N_\star\gtrsim 100$, i.e., the posterior distributions are approximately Gaussian.
    For $N_\star\lesssim 100$, non-Gaussianity may need to be considered, although the choice of priors likely has a stronger influence on the final results (see Section~\ref{ssub:priors}).
    \cite{Joachimi2011-er} have shown it is possible to correct Fisher analysis results in order to approximate a non-Gaussian posterior and we leave this for future work.

\section{Discussion and conclusions}\label{conclusions}
    We have studied the prospects for constraining the central densities of dark matter in dSphs for current and future data sets.
    Our Fisher matrix framework allows us to quickly forecast uncertainties on density profile parameters, using individual stars, without running a full computational-intensive Jeans Equation analysis.
    The main results of the paper are as follows:
    \begin{enumerate}
        \item Considering the errors on all our parameters, we demonstrate that the enclosed mass has a minimum uncertainty at $r_{1/2}$.
        Previous work \citep{Wolf2010-xe,Walker2009-fv,Lazar2020-jr} showed this analytically only considering the stellar anisotropy.

        \item For a Draco-like system ($\sigma_{\mathrm{los}}\sim10~\kms$), improving current RVs precisions ($\delta v = 2~\kms$) will not yield significantly better results on density parameter estimates.
        
        \item For a Draco-like system, 1000 radial velocity measurements can constrain the enclosed mass to 5\%, however proper motions are needed to place similar constraints on the shape of the density profile.
        
        \item For systems with fewer than 100 stars, the choice of priors completely determines the resulting density profile parameter constraints, as there is not sufficient information to constrain these parameters.
        In this regime, informative priors, e.g. from simulations, may be useful.

        \item We considered two systems, one with a cusp density profile, the other a cored density profile, which resulted in similar velocity dispersions.
        By analyzing the 2D confidence region between the inner-slope and stellar anisotropy we determined that $\sim100000$ stars with full 3D velocity information are required to distinguish between these two systems at $2\sigma$ level.

        \item The log-slope measured at $r_{1/2}$ is easier to constrain only requiring around 1000 stars to make a $2\sigma$ detection.
        While the log-slope is well constrained at this radius, whether this measurement is informative depends on the nature of the density profile. For example, a measurement of the log-slope at this radius would be able to rule out a large core, however, it would be much less sensitive to a smaller core.
    \end{enumerate} 

    Though we have chosen to perform our analysis independent of any specific dark matter model, e.g., self-interacting or warm dark matter, our conclusions may be strengthened by appealing to the predictions of these models.
    In particular, simulations in these different models predict different relationships between the maximum circular velocity and radius of maximum circular velocity, or equivalently the scale density and the scale radius, for subhalos.
    Ultimately, these relationships may be implemented as priors in either a Fisher matrix or likelihood scanning formalism, and a comparison of the behavior of the different models with both line-of-sight velocities and proper motions may strengthen the dark matter constraints that are obtained.

    In using both radial velocity and proper motion measurements of individual stars, we have exploited information in five out of six possible phase space coordinates.
    The one piece of information that we have not included is the distance to the member stars in the dSphs.
    While individual stars can be identified as having a distance consistent with being members of a given dSph, information on their internal position are not available.
    However, for stars such as RR Lyrae, which do have individual measured distances~\citep{Richardson2014-wr,Ferguson2020-pi}, it may be possible to use this information in the future, particularly if the uncertainties can be reduced to $\lesssim 5$\%~\citep{Sesar2018-ln}.
    With this information, the 3D positions of stars can be taken into account in our likelihood function, and may allow for better constrain our model parameters.
    In future modeling it will be interesting to see if this distance information provides stronger constraints within the context of our likelihood analysis.

    For future observations, systems with velocity dispersions of $\sim 10$\,\kms (e.g., Draco, Sculptor, Fornax) do not require velocity precision higher than $\delta v = 2~\kms$ in order to improve our knowledge of these system.
    In these cases, the number of stars will play a significantly more important role.
    Additional constraining power will also result from improved membership probabilities, stellar distances, multiple stellar populations, chemical abundances, etc.
    A number of current and near-term facilities are posed to dramatically improve available kinematic data sets for Milky Way satellite systems.
    The methodology presented here can help optimize use of these facilities to improve density profile uncertainties in systems with resolved star kinematic data.

\medskip

The authors thank James Bullock, Manoj Kaplinghat, Ethan Nadler, Lina Necib, Imad Pasha, Justin Read, Nathan Sanford and Josh Simon for helpful discussions and feedback that have improved this manuscript.
MG acknowledges support from the Howard Hughes Medical Institute through the Professors Program.
LES acknowledges support from DOE Grant de-sc0010813.
This work was supported in part by a Development Fellowship from the Texas A$\&$M University System National Laboratories Office.

\bibliography{ref.bib}
\end{document}